# From Rome to the Antipodes: the medieval form of the world

Amelia Carolina Sparavigna
Department of Applied Science and Technology, Politecnico di Torino, Italy

*Here we discuss how some medieval scholars in the Western Europe viewed the form of the world and the problem of the Antipodes, starting from the Natural History written by Pliny and ending in the Hell of Dante Alighieri. From the center of the Earth, Dante and Virgil came to the Antipodes: eventually, their existence was accepted. Among the others, we will discuss the works of the Venerable Bede and Sylvester the Second.*

The most commonly given date for the beginning of the Middle Ages is 476 AD, when the emperor Romulus Augustus, of the western part of the Roman Empire, was deposed [1]. The end of the Middle Ages is coincident, depending on the popular context, on events such as the Christopher Columbus's first voyage in 1492. The crisis of the Western Roman Empire in fact, was started a century before, when a large numbers of Goths crossed the Danube River, the Eastern border of the Empire. The deposition of the last roman emperor had as a consequence that the empire was scattered among the conquerors.

After a century from their coming, Theodoric the Great (454-526 AD) became king of the Ostrogoths and regent of the Visigoths, sovereign of a large part of the western roman territories. It was a period of religious struggle among Christians and against Pagans. Within his kingdom, Theodoric allowed Roman citizens to be under the Roman law. The Goths, meanwhile, lived under their own customs [2]. Theodoric was of Arian faith: at the end of his reign some quarrels arose with the Byzantine emperor and the relations deteriorated. After Theodoric's death, the kingdom of the Ostrogoths began to wane and was conquered by the Byzantine Empire. We have to wait until the kingdom of Charlemagne (800 AD) to see the rise of a Roman Empire again.

From the fourth to the sixth century then we can see the decline of the Roman civilization. Besides the dissolution of the Roman infrastructures, we have also that the ancient Greek knowledge and Roman culture began to fade away. A reason for this lost of civilization was a scarce knowledge of the conquerors of the cultures of ancient Asia and Greece. Moreover, there was the problem of the new Christian religion and of what is told in the Bible about the world. As explained by J.L.E. Dreyer in [3], the conquerors of the Western Roman Empire gradually adopted the Christian religion, "but most of their teachers were unfortunately devoid of sympathy for anything that emanated from the heathen Greek and Roman world; and it was left to the dying Neo-Platonic school and to pagan commentators like Macrobius and Simplicius and the encyclopedic writer Martianus Capella to keep alive for a while the traditions of the past."

There are several subjects of discussion then, about the dissolution of the ancient culture after the end of the Western Roman Empire. Let us choose one of the problems that the ancient Greece and Rome had investigated, that is, that of the shape of the world and the existence of Antipodes. As a starting point we have to consider what Pliny the Elder is writing in his Natural History [4], because his book was one of the encyclopedic works of the Latin culture that survived in the Middle Age. Then, we start before the medieval times, from the Rome of the first century AD. Our discussion will end with Dante Alighieri and his Hell at the center of Earth. From the center of the Earth, Dante and Virgil climb to the Antipodes, to "riveder le stelle", to see the stars again. To have the science of Galileo and Newton, people still had to climb a high "Purgatory Mountain".

## 1. Pliny and the spherical symmetry

As we have discussed in some previous papers [5,6], Pliny reported quite interesting descriptions of physical phenomena. Pliny discussed a sort of "gravitation" too.

For him, the world (mundus), that is the heaven, cosmos or universe, was a globe. "That it has the form of a perfect globe we learn from the name which has been uniformly given to it, as well as from numerous natural arguments. For not only does a figure of this kind return everywhere into itself and sustain itself, also including itself, requiring no adjustments, not sensible of either end or beginning in any of its parts, and is best fitted for that motion, with which, as will appear hereafter, it is continually turning round; but still more, because we perceive it, by the evidence of the sight, to be, in every part, convex and central, which could not be the case were it of any other figure." John Bostock [7], that translated and commented Pliny, remarked that Pliny's astronomy for the most part, derived from Aristotle. The fact that the figure returns into itself, that is the spherical symmetry, is coming from Plato's Timæus, and adopted by Aristotle, in De Cœlo, and by Cicero. The spherical form of the world, ouranos, and its circular motion reappears in the Magna Constructio by Ptholemy, frequently referred to by its Arabic title Almagestum (125-140 AD). Ptholemy's astronomical work was translated into Arabic in 827 AD, and then, came back in the Western Europe in a Latin translation.

And what was the shape of our planet according to Pliny? From the Chapter 64, Of the Form of the Earth, "Every one agrees that it has the most perfect figure. We always speak of the ball of the earth, and we admit it to be a globe bounded by the poles. It has not indeed the form of an absolute sphere, from the number of lofty mountains and flat plains; but if the termination of the lines be bounded by a curve, this would compose a perfect sphere. And this we learn from arguments drawn from the nature of things, although not from the same considerations which we made use of with respect to the heavens. For in these the hollow convexity everywhere bends on itself, and leans upon the earth as its centre. Whereas the earth rises up solid and dense, like something that swells up and is protruded outwards. The heavens bend towards the centre, while the earth goes from the centre, the continual rolling of the heavens about it forcing its immense globe into the form of a sphere". A mechanical analogy with rolling spheres is used to describe this world turning about the Earth, that is, about its axis which is the Axis Mundi, the axis of the cosmos.

However, if our planet is a sphere, another question arises, in the Chapter 65, Whether there be some Antipodes? "On this point - Pliny tells - there is a great contest between the learned and the vulgar. We maintain, that there are men dispersed over every part of the earth, that they stand with their feet turned towards each other, that the vault of the heavens appears alike to all of them, and that they, all of them, appear to tread equally on the middle of the earth. If any one should ask, why those situated opposite to us do not fall, we directly ask in return, whether those on the opposite side do not wonder that we do not fall". Good idea the use of symmetry.

And Pliny continues discussing the surface of water in oceans, which seems our discussions on the necessity its surface be equipotential. "But what the vulgar most strenuously contend against is, to be compelled to believe that the water is forced into a rounded figure; yet there is nothing more obvious to the sight among the phenomena of nature... It is from the same cause that the land is not visible from the body of a ship when it may be seen from the mast; and that when a vessel is receding, if any bright object be fixed to the mast, it seems gradually to descend and finally to become invisible. And the ocean, which we admit to be without limits, if it had any other figure, could it cohere and exist without falling, there being no external margin to contain it? ... For since water always runs from a higher to a lower level, and this is admitted to be essential to it, no one ever doubted that the water would accumulate on any shore, as much as its slope would allow it. It is also certain, that the lower anything is, so much the nearer is it to the centre ... Hence it follows, that all the water, from every part, tends towards the centre, and, because it has this tendency, does not fall."

In Pliny then, we find the popular argument showing the earth as a sphere, using a ship over the horizon.  Pliny also discusses what regulates the days and nights. "Hence it is that there is not any one night and day the same, in all parts of the earth, at the same time; the intervention of the globe producing night, and its turning round producing day. This is known by various observations. In Africa and in Spain it is made evident by the Towers of Hannibal, and in Asia by the beacons ... for it has been frequently observed, that the signals, which were lighted at the sixth hour of the day, were seen at the third hour of the night by those who were the most remote."

We see that Pliny's Mundus was a geocentric sphere. Let us remember that a heliocentric system was proposed by Aristarchus of Samos. This scientist lived in the 3rd century BCE. He measure the distance of the Moon using lunar total eclipses. In opposition to the common belief of a geocentric universe, he told that the Earth revolves around the Sun, besides rotating on its axis. The original Aristarchus' work on the motion of Earth had not survived, but his ideas are known from the discussion on them by Archimedes, in his treatise "The Sand-Reckoner". Archimedes is telling that Aristarchus proposed that the Universe was vastly larger than was commonly believed [8].

**2. The Bible and the water above the firmament**

As we have previously told, the ancient Greek and Roman culture was not appreciated by some Christians. As it is told in [3], even before enemies from outside had begun to assail the Roman Empire, "a narrow-minded literal interpretation of every syllable in the Scriptures was insisted on by the leaders of the Church, and anything which could not be reconciled therewith was rejected with horror and scorn." And then the hand of time had been wrenched back of a thousand years, and centuries were to elapse before science recovered [3]. Emblematic is the death of Hypatia on a day of March 415 [9]. She was a Greek philosopher and mathematician in the Roman Egypt.  As head of the Platonist school at Alexandria, she also taught philosophy and astronomy, encouraging logic and mathematics. Hypatia was murdered by Christians after being accused of exacerbating a conflict between the governor and the bishop of Alexandria.

Reference [3] tells us that there was also a precise desire to sweep away all the results of Greek learning regarding the figure of the Earth and the motion of the planets. However, at first, some of the followers of the Apostles had no problems with the Greek science. Clemens Romanus, in an epistle to the Corinthians, written about 96 AD, alludes to the Antipodes as a part of the Earth to which nobody can approach, and uses a typical expression of the classical writings, that "the sun and moon and the dancing stars, according to God's appointment circle in harmony within the bounds assigned to them without any swerving aside." [3].

In Alexandria of Egypt, people had more familiarity with the philosophical speculations of the Neo-Platonists. Therefore, Clement of Alexandria (about 200 AD) was the first to view in the Bible some descriptions as representing allegorically the world. [3,10]. This desire to find allegories in Scripture was followed by Origen (185-254 AD) [11]. To avoid problems with the Greek science, he discussed, for instance, the separation of the waters above the firmament from those below it [12], that we find in Genesis of the Bible, considering it as an allegory meaning that we have to separate our soul from the darkness of the Adversary. "But this kind of teaching was not to the taste of those who would have nothing to do with anything that came from the pre-Christian world" [3]. Therefore, there were men as Lactantius [13], that was the first and the worst of the adversaries of the rotundity of the earth. In one of his books, written between 302 and 323 AD, "On the false wisdom of the philosophers", a chapter is a mockery of the idea of a spherical figures of Earth and the existence of antipodes [14]. Lactantius considered the argument of philosophers that heavy bodies fall towards the centre of Earth as unworthy of serious notice.

Basil of Cesarea [15], called the Great, was not afraid by the opinions of philosophers as Lactantius did [3].  Basil is writing for instance "We say that the universe is one in number, but not that it is one in nature, nor yet that it is simple; for we divide it into the elements of which it is composed: fire, water, air, and earth. Again, man is called one in number; for we often speak of one man. But he is not simple, since he is composed of body and soul [16]".

Here we find the four elements of the Greek world. For what concerns astronomy, he knew the fact that there are stars about the South Pole which we cannot see at northern latitudes, and he understood how summer and winter depend on the motion of the sun through the zodiac [3]. However, the Bible tells that above the firmament waters exist, and then he proposed that these waters were placed above there to prevent the world from being burned by the celestial fire. As to the figure of the earth, he says that many have disputed whether the earth is a sphere or a cylinder or a disc, or whether it is hollow in the middle [3].

The literal interpretation of the Bible was totally followed by the leaders of the Syrian Church, who accepted only the cosmogony of the Genesis. Some contemporaries of Basil, Cyril of Jerusalem and Severian of Gabala agreed with the creation of the world according the Genesis. The heaven is not a sphere, but a tent, a tabernacle, a vault, or a curtain. The earth is flat and the sun does not pass under it in the night, but travels through the northern parts, hidden by a wall. Diodorus, Bishop of Tarsus (died 394), was against those atheists who believe in the geocentric system. And St Jerome wrote violently "against those who followed the stupid wisdom of the philosophers" [3].

Ambrose of Milan (d. 397 AD) had no interest in the nature of the earth, or whether the world is made of the four elements or of a fifth, however, for him the heaven was a sphere. Here we see that he knew Aristotle that was the ancient philosopher adding a fifth element, the "aether", to the classic fours. This fifth element became the "quintessence". Augustine (354-430), a disciple of Ambrose, with regard to the Antipodes, says that there is no historical evidence of their existence, and, following the same reasons of Pliny, tells that some people guess that the opposite side of the earth, which is suspended in the convexity of the world, cannot be devoid of inhabitants. However, he consider absurd to imagine that "people from our parts could have navigated over the immense ocean to the other side, or that people over there could have sprung from Adam". As told in [3], none of these Fathers of the Church decided to condemn the Greek astronomy or undertaken to work on a system able to substitute the doctrines of the pagan philosophers.

**3. Cosmas, the monk**
It could be surprising, but such a work was undertaken by a person that travelled by lands and by seas: Cosmas the Indicopleustes, or the Indian navigator. His book, the Christian Topography [17], had been written between the years 535 and 547. The titles of the chapters are quite clear: I - Against those who want to be Christian, but believe and profess, like those outside, that heaven is spherical, II - Christian theories on the form and layout of places in the whole universe, taking their proofs from the divine scriptures, III - That divine scripture is sure and worthy of trust, that it reveals things which agree among themselves and with the whole, in both the Old and New Testament, and that it indicates the utility of the shape of the whole universe, IV - Concise recapitulation, with illustration, of the shape of the universe according to the divine scriptures, and refutation of the sphere, V - Where is found the description of the tabernacle and the agreement of the prophets and the apostles, and do on.

Cosmas was probably a native of Alexandria, and during the earlier part of his life he was a merchant. He travelled in the Mediterranean, the Red Sea, and the Persian Gulf, and on one occasion he even dared to sail on the Ocean. As [3] is telling, he "describes his travels in Abyssinia and adjoining countries. As he must have reached places within ten degrees of the equator, it is very remarkable that he could be blind to the fact that the earth is a sphere."

According to Cosmas, the position of the Earth is not at the centre of the universe, because it is so heavy that it can only find rest at the bottom of the universe. He is telling "even in supposing that the earth is in the middle of the universe, as its centre, you immediately give the deathblow to your own theory when you repeat that the middle is below, for it is impossible that the same thing can at once both be in the middle and below, for the middle is the middle between up and down, or between right and left, or between before and behind. Why do you then, when beleaguered with difficulties, utter absurdities contrary to nature, in opposition to scripture?"
Any concept of symmetry is completely lost.
Elaborating the question of the Antipodes, according to Cosmas, one "would easily find them to be old wives' fables. For if two men on opposite sides placed the soles of their feet each against each, whether they chose to stand on earth, or water, or air, or fire, or any other kind of body, how could both be found standing upright? The one would assuredly be found in the natural upright position, and the other, contrary to nature, head downward" [17]. The only relevant sentence in his book is that on the "old wives' fables": who is writing (ACS) is accordingly supposing that at Cosma's times, women knew the Greek philosophy better than monks did.
Cosmas thinks that the figure of the universe can only be learned by studying the design of the Tabernacle of Moses [18,19]. The table of shew-bread with its wavy border signified the earth surrounded by the ocean. As the table was positioned from East to West, we have that the earth is a rectangular plane, twice as long as broad, where the longer dimension is going from East to West. And so, we find a flat earth. In fact, this view of the earth and cosmos probably was not coming from Cosma's mind, but taught by some Churches of the preceding two hundred years [3], because Cosmas is collecting several quotations from the Fathers of Church, in particular from Severianus. As Dreyer concludes in his book, some persons used the "authority of their position in the Church and their literary ability to propagate ideas which had been abandoned in Greece eight hundred years before. But what Kosmas does deserve to be blamed for is his not finding out on his travels that the earth is a sphere." [3] And we can add that it is so easy, as Pliny tells, to observe a ship over the horizon of the sea.
However, some writers continued to study the works of the Greek philosophers. Among these was Johannes Philoponus, a grammarian of Alexandria, who wrote commentaries to the writings of Aristotle, showing a remarkable freedom of thought. As a result, he became a "heretic". Then, there was Isidorus Hispalensis, Bishop of Seville in 601. Known by his erudition and eloquence, he wrote an encyclopedic work, "Etymologiarum libri", in the beginning of which he enumerates the seven free arts, grammar, rhetoric, dialectics (the trivium), and arithmetic, music, geometry, astronomy (the quadrivium). In this work, Isidore explains the meaning of a word and the origin of an expression. According to [3], when dealing some topics, such as the form of the world and the earth, "he does not lay down the law himself, but quotes "the philosophers" as teaching this or that, though without finding fault with them. Thus he repeatedly mentions that according to them the heaven is a sphere, rotating round an axis and having the earth in the centre."
However, though some scholars like Philoponus and Isidore had accepted some of the ancient science, there were also cosmographers like Cosmas. Among them, Dreyer cites a geographer of Ravenna. "The earth of course is flat, the sun likewise (it is spoken of as a table, mensa solis), and it passes through the gate of the east every morning to lighten up the world, and passes in the evening through the gate of the west to return during the night to its starting-point through the south. … No doubt there continued throughout the Middle Ages to be clerics to whom the sphericity of the earth was an abomination … But in the peaceful retreat of the monastery the study of the ancient Latin writers had long before the time of the Ravennese geographer taken root, and the geocentric system slowly but steadily began to resume its place among generally accepted facts." [3]

**4. The hours of Bede the Venerable**
Bede was born, about 673 AD, in the north of England. He spent most of his life in two monasteries of that region, where he had to possibility to read a large number of books, brought from Rome by their founders. His reputation increased after his death so largely that he got the title of "Venerable". And therefore we have some spurious treatises among the genuine ones. A genuine writing of Bede is a treatise entitled "De natura rerum", where he discusses the Earth and its divisions, thunder, earthquakes, and so on. In [3], it is remarked that the contents of "De natura rerum" are taken from Pliny, often almost verbatim. Therefore we have not to be surprised that for Bede, the Earth is a sphere and the seven planets are moving round it. The sun is rotating about the Earth and so on.

In his Ecclesiastical History of England [20], in the chapter "Of the Situation of Britain and Ireland, and of their ancient inhabitants," he tells that "because it lies almost under the North Pole, the nights are light in summer, so that at midnight the beholders are often in doubt whether the evening twilight still continues, or that of the morning has come; since the sun at night returns to the east in the northern regions without passing far beneath the earth. For this reason the days are of a great length in summer, and on the other hand, the nights in winter are eighteen hours long, for the sun then withdraws into southern parts. In like manner the nights are very short in summer, and the days in winter, that is, only six equinoctial hours. Whereas, in Armenia, Macedonia, Italy, and other countries of the same latitude, the longest day or night extends but to fifteen hours, and the shortest to nine." [20] It is interesting to note that he uses the "equinoctial hours", when we have twelve hours for the day and twelve for the night.

If we consider an hour as one twelfth of the time from sunrise to sunset, according to the latitude, the hours on summer days were longer than on winter days. Then, these hours are sometimes called temporal, seasonal, or unequal hours and Romans used this definition. Also the night was divided into twelve hours.

Let us assume to be on an equinox, if we consider the hour as one twenty-fourth of this day, this hour, which is the hour used by Bede, is equivalent to the hour defined as one twenty-fourth of the mean solar day. Bede had a unit of measurement and compared the length of days and nights at different latitudes. We can guess another fact, that Bede had probably an equatorial sundial. This instrument, also called equinoctial sundial, is composed by a planar surface and a corresponding perpendicular gnomon. The plane surface is equatorial, because it is parallel to the equator of the Earth and celestial sphere. If the gnomon is fixed perpendicularly, it is parallel to the axis mundi. In this sundial, the hour-lines are all spaced 15° apart (360°/24).

May be, Bede had an armillary sphere too. The armillary sphere dates from about 225 BCE and was created by Eratosthenes. The name armillary comes from the Latin armilla, bracelet. In this sundial we have concentric rings which represent the Equator, the Horizon and the Meridian, and the Gnomon is the axis mundi.

The fact that the length of the day depends on the latitude is natural when the earth is a globe, but there were people believing in a flat earth and for them the length of the days was the same all over the world.

Bede wrote another book, De temporum ratione, that is a book on the subdivision of time of the year. From the patristic age until the Gregorian calendar reform of 1582, the computus, that is the time reckoning to have a calendar, was quite important. The book written by Bede was the first comprehensive treatise on this subject. He describes the Julian solar calendar and the tables and formulae for calculating dates [21].

**5. Fergil the Irish and the "heresy" of the Antipodes**
Fergil or Virgil was an Irish ecclesiastic of the eighth century, better known as Virgilius of Salzburg. He was originally Abbot of Aghaboe. He started a pilgrimage to the Holy Land about 745, but he stopped at Salzburg, where he became Abbot of St Peter's. In 748 he had problems with Boniface, the head of the missionary Churches of Germany. Boniface reported to the Pope Zacharias, that Virgil in his lectures had taught that there was "another world and other people under the earth." Zacharias replied that Boniface should call a council and expel Virgil from the Church, if he really had taught that. Virgil was not persecuted because he became Bishop of Salzburg in 767, however, when both Boniface and Zacharias were long dead [3]. None of his writings and nothing of his doctrine survived. We know just the words from the Pope's reply, about the "heresy" of Virgil on the existence of antipodes. As we will see in the last section of this paper, another Virgil, the ancient Latin poet, had been chosen by Dante to lead him through the earth till the Antipodes, in his Divine Comedy.

According to [3], "after all there is nothing very remarkable in the fact that an Irish monk knew the earth to be a sphere. Not only were many Irish monasteries centres of culture and learning, where the fine arts and classical literature were studied at a time when thick night covered most of the Continent and to a less extent England; but devoted missionaries had before the time of Virgil spread the light of Christianity as far north as the Orkneys". The sphericity of the earth was for them an undoubted fact [3].

In the following century, another Irishman, Dicuil, wrote a geographical compilation, the "Liber de mensura orbis terrae", in 825 [22,23]. This book is providing information about various lands. This work was based upon a "Mensuratio orbis" prepared by order of Theodosius II (435 AD), copied by the Carolingian court, and on the texts written by Pliny the Elder and others, among them, Isidore of Seville. In any case, he adds the results of his own investigations [22], and a few reports which he got from the travelers of his time [3]. For instance, from a monk, Fidelis, he is reporting of a canal existing between the Nile and the Red Sea. From some clerics, he writes information about the Faroe Islands and Iceland.

He tells that the Irish missionaries who visited Thule, our Iceland, saw the sun at midnight in midsummer, "It is now thirty years since certain priests, who had been on that island from the 1st of February to the 1st of August, told that not only at the time of the summer solstice, but also during the days before and after, the setting sun at evening conceals itself as it were behind a little mound, so that it does not grow dark even for the shortest space of time, but whatsoever work a man will do, even picking the lice out of his shirt, he may do it just as though the sun were there, and if they had been upon the high mountains of the island perhaps the sun would never be concealed by them [i.e., the mountains]. In the middle of this very short time it is midnight in the middle of the earth, and on the other hand I suppose in the same way that at the winter solstice and for a few days on either side of it the dawn is seen for a very short time in Thule, when it is midday in the middle of the earth. Consequently I believe that they lie and are in error who wrote that there was a stiffened (concretum) sea around it, and likewise those who said that there was continuous day without night from the vernal equinox till the autumnal equinox, and conversely continuous night from the autumnal equinox till the vernal, since those who sailed thither reached it in the natural time for great cold, and while they were there always had day and night alternately except at the time of the summer solstice; but a day's sail northward from it they found the frozen sea." [24]

We have no doubts that Dicuil saw the earth as an oblique globe.

**6. Sylvester II, astronomer and pope.**

In 999, a mathematician, Gerbert of Aurillac, ascended the papal throne as Sylvester II. He was familiar with the scientific writings of ancients. Before being a pope, he constructed astrolabes, and celestial and terrestrial globes to assist his lectures on astronomy [25].

Gerbert was born in the region of Auvergne, in central France. About 963, he entered the monastery of St. Gerald at Aurillac that Gerald the Good had established near his castle some sixty years earlier. Here he studied the Latin trivium, that is, grammar, logic, and rhetoric. It happened that in 967, a noble man from Barcelona visited the monastery, and the abbot asked him to take Gerbert back to Spain so that Gerbert could study mathematics there, that is, to study the quadrivium: arithmetic, geometry, music, and astronomy [25]. In Spain, Gerbert was put in the care of the bishop of Vic, in Catalunya. This region was a frontier territory, where there was a considerable cultural exchange with the Muslims of al-Andalus.

Al-Andalus was much more advanced that Christian Europe. For comparison, Ref.[25] tells that while the greatest library in Christian Europe had less than a thousand volumes, the library of Cordoba had over four hundred thousand of books. Due to their proximity to Al-Andalus, the libraries of Vic and of the nearby monastery of Ripoll were among the largest in Europe. At those times, the Muslim astronomy was the most advanced in the world, and Muslim astronomers using the astrolabe had prepared the best maps of the skies [25]. Let us note that the names of main stars are Arabic.

The Arabs were even further advanced in arithmetic. They knew the zero from the Indians and used a positional numeric system much like the modern system (in Europe, at those times the Roman system was used). But they also established the algebra, investigating prime numbers and equations. Since they studied the proportions, it was possible for them "to approach music in a quite precise manner, distinguishing accurately between notes, developing theories of harmonies and discords, and constructing musical instruments with quite accurate tuning" [25]. For this reason, at the school of the Vic cathedral, Gerbert was able to learn much of this knowledge. And Gerbert was so clever to take full advantage of this opportunity.

In 969, Gerbet was pilgrim in Rome, where he met the Pope John XIII (965-971) and the emperor Otto I (962-973). The pope persuaded Otto to have Gerbert as tutor for his young son; after some years, Gerbert went to study advanced logic at the school of Reims. There, he built an organ with constant pressure supplied by water power, instead of using the pressure generated by the organist or an assistant pumping some bellows. The Gerbert's organ had a steady level of sound and pipes that were matched mathematically. He also studied the abacus, even constructing a giant one [25].

When his pupil Otto became Holy Roman Emperor in 983, he made Gerbert the abbot of the monastery of Bobbio, founded by St. Columban, and having one of the greatest libraries in the Western Europe. Otto died the next year, however, and Gerbert returned to Reims as the master of the cathedral school and secretary to the archbishop. In the following years, he became deeply involved in the political struggles of the times (for all the details, please see [25]). Eventually, Gerbert was called to become the teacher and advisor of Otto III, in Ravenna, the southern capital of the Holy Roman Emperors. When Pope Gregory V died in 999, Otto proposed Gerbert as pope. He took the name Sylvester the Second. As remarked in [25], Sylvester the First was the advisor of the emperor Constantine. But the Roman people rebelled against a foreign pope, and both Otto and Gerbert were forced to flee to Ravenna. Otto died in 1002, and Sylvester II returned to Rome but died a little later.

Gerbert is credited to have reintroduced the astronomical armillary sphere to Latin Europe via Al-Andalus in the late 10th century [3], the details of these spheres are revealed in his letters. However, there is the possibility that in some places of Europe, the use of armillary spheres had not ceased (see the discussion in section 4). There was another sphere that Gerbert used as a visual aid for teaching astronomy in the classroom: let us define it the Gerbert's sphere. It was a sphere suitable for his astronomical observations, equipped with sighting tubes [26] and a graduated scale having 60 grades, instead of 360 [27].

With Sylvester II, we can tell that the rotundity of the earth and the geocentric system of planetary motions were reinstated in the human knowledge, with the same role they had held "among the philosophers of Greece from the days of Plato" [3]. Unfortunately, the original works of these ancient philosophers were not read, because Greek had been an unknown tongue after the fifth century. However, as in the case of the Venerable Bede, the Latin writers and among them Pliny were a source of information to anyone who read them, and since the days of Charles the Great (768-814) Roman literature was rapidly becoming better known [3,28].

**7. Imago Mundi**

Honorius Augustodunensis (1080-1154), known as Honorius of Autun, was a very popular theologian. He wrote with a lively style, approachable for the common people. About his life, he was a monk and that he travelled to England and was a student of Anselmo d'Aosta. Among his works we have the "Imago mundi", which contained cosmology and geography and some chronicles. Since this book was translated from Latin in vernacular languages, it became popular throughout the medieval period [29]. The parts of the book on the cosmos were deduced from the work of Pliny. In agreement with a medieval terminology, the upper heaven is called "firmament". The firmament is a sphere adorned with stars and outside this sphere there in the water, mentioned by the Bible, in the form of clouds. Above these clouds there is the spiritual heaven, where the angels live on nine orders. Here is the Paradise. God created the heaven and the earth. The earth is a round like a ball or like an egg, with yolk, white and shell around it (est enim in perpetuo motu, huius figura est in modum pile rotunda sed instar ovi elementis distincta) [30,31].

The encyclopedic works, such as the Imago Mundi, had a strong development in the XIII century. Moreover, such works were influenced by the Aristotle's philosophy, which had begun to circulate in France in Arabic translation, introduced from Spain [3,25]. Neo-platonic and some Arabian works, masked as commentaries of Aristotle's works, were also read. As a consequence, during a council held at Paris in 1209, it was decided that it was forbidden to read in public or privately the Aristotle's books on Natural Philosophy and commentaries on them [3]. In 1215 this prohibition was renewed in the University of Paris. However, in 1254 the situation was completely different, and the University stated how many hours should be used in teaching Aristotle and his natural philosophy. From Paris, Aristotle spread in Europe. After the translations from Arabic to Latin, translations made directly from Greek were provided by Albertus Magnus (1193-1280) and Thomas Aquinas (1227-1274). In the thirteenth century, Aristotle became the ancient philosopher recognized as the best resource for theologians.

**8. Grosseteste and the sphere of light**
Robert Grosseteste (1175-1253) was an English philosopher who became the Bishop of Lincoln. As a scientist he had a quite important role in the medieval school of Oxford [32]. In his works, in particular in the commentaries of Aristotle's philosophy, Grosseteste devised a scientific method. From particular observations, we can find a universal law; from these laws we can predict some particular cases. Grosseteste called this "resolution and composition" [33]. As a consequence, Grosseteste needs to verify the physical laws through experimentation. These ideas were a prelude for the Galilean science in the 17th century [34].

The method of "resolution and composition" was applied to geometry and optics. However, optics is described by geometry, because optics depends on geometry. As a conclusion, Grosseteste argued that mathematics was the highest science, basis for all others. Here we see that he understood the necessity to describe the physical phenomena in a mathematic formalism. Since the light is the visualization of geometry, in the geometrical optics, Grosseteste believed that at the beginning, it was the light to move the universe.

In his "De Luce", Grosseteste explains the origin of the world. God created matter and light together in a point. Due to its nature the light propagated isotropically in all directions. It immediately became a sphere and, accordingly, dragged by the light, the matter started to expand. The creation is then explained by means of a sphere of light [34].

Grosseteste's work in optics was continued by Roger Bacon. There is also an interesting quotation often reported in the history of telescope. In "De Iride", Grosseteste writes that a part of optics, "when well understood, shows us how we may make things a very long distance off appear as if placed very close, and large near things appear very small, and how we may make small things placed at a distance appear any size we want, so that it may be possible for us to read the smallest letters at incredible distances, or to count sand, or seed, or any sort of minute objects." He probably had an optical bench.

**9. Tommaso d'Aquino and Aristotle**
The most representative philosopher among the scholastics is Tommaso d'Aquino, Thomas Aquinas (1225 –1274). He was an Italian Dominican priest. Due to his influence in the tradition of scholasticism, he is also known as an Angelic Doctor or Doctor Universalis. Thomas came from a noble family of the Kingdom of Naples, with the title of counts of Aquino. His influence on Western thought is considerable, because some of the following philosophy was developed in continuing or refuting his ideas about ethics, natural law, and metaphysics.

His best-known works are the "Summa theologiae" and the "Summa Contra Gentiles". He also wrote a commentary on the book on the heavens of Aristotle. In discussing the Aristotle's cosmology [35], Thomas was never confused by the relevant differences between the opinions of the ancient philosopher and the content of the Bible [3]. In his commentary [35], Thomas quotes that written by Simplicius [36], Plato, Ptolemy, and others. In discussing the position of the earth at the centre of the world, he uses the Ptolemy's arguments.

In "the Heavens" [37], Thomas Aquinas describes the motions of bodies. We find that "straight motion naturally belongs to someone or other of the simple bodies, as fire is moved upward and earth downward and toward the middle. And if it happens that a straight motion is found in mixed bodies, that will be due to the nature of the simple body predominant in it." As Aristotle believed, the cause of the downward motion of heavy bodies toward the center of the world was in the nature of the bodies. They were made by the earth element. Light bodies, such as those made by the element fire, move upward toward the sphere of the Moon. Let us remember, that Aristotle had the four elements, corruptible, which were moving naturally on straight lines. If we do not see a straight motion, it means that it is not natural but a violent one, such as that of a stone launched upwards. In any case, these motions had a beginning and an end. For this reason, they were not perfect. But in the sky, we see the circular motions, and the circular motion is perfect, having no beginning or end. The element having a natural circular motion is the "aether", an incorruptible element. After this short discussion on the motion as viewed by Aristotle, let us remark that both Aristotle and Thomas did not know the principle of inertia [34].

According to [3], another book was quoted in the textbooks on astronomy for nearly four centuries. His author was Johannes del Sacrobosco, or John of Holywood, who died at Paris in 1256. The book is entitled "De sphaera mundi", On the Sphere of the World and is an introduction to the elements of astronomy, based on Ptolemy's Almagest. Some additional ideas were coming from the Islamic astronomy. Johannes quotes Ptolemy and Al-Farghani [3,38]. He was then the first European writer to give a sketch of the Ptolemaic system of planetary motions. And then Ptolemy began to be the master, because the Sacrobosco's De sphaera mundi was used in universities for hundreds of years and the manuscript copied many times before the invention of the printing press. The first printed edition appeared in 1472 in Ferrara, followed by least 84 editions in the next two centuries. The number of edition reflects its importance as a university text [39].

However, according to [3], a large part of the scholastics considered that the study of Aristotle's works was enough. "One man there was, however, who was not content to be a mere slave of Aristotle, any more than the Alexandrian thinkers had been". Roger Bacon (1214-1294) in his Opus Majus shows that he knew the literature of the Greeks and Arabians, but "he is able to think for himself, and he lays stress on the importance of experiments as offering the only chance of helping science out of the state of infancy in which he is fully aware it still lies." [3]

**10. Roger Bacon**

Roger Bacon, (1214–1294) was an English philosopher and a Franciscan friar, and a scientist because he emphasized the study of nature through empirical methods. According to him, the light is "forma prima". All the relationships among bodies are explained using the same laws ruling the propagation of light. The action-reaction between bodies is given by lines of force, straight because of a principle of least action. The speed of light is finite [24]. In reading [3], we discover that the environment was not positive to Bacon and that its works "had to lie in manuscript for nearly five hundred years before it was printed ".

However, about the universe, he followed Ptolemy. He tells that the earth is an insignificant body at the centre of the vast universe, and following Al-Farghani, even the smallest star is larger than the earth. Moreover, he tells that Ptolemy showed that a star takes 36,000 years to travel in the sky (that is, Ptolemy knew the precession, although not its precise value), while a man can walk round the earth in less than three years [3]. He discusses also the question concerning the surface of the earth covered by the sea. Using some statements of Aristotle, Seneca, and Ptolemy, he concludes that the ocean between the East coast of Asia and the Europe is not so large [3]. This conclusion was almost literally copied by the Cardinal d'Ailly in his Imago Mundi (written in 1410 and printed in 1490), without any mention of Bacon [3]. After, Columbus reported this in a letter to the Spanish monarchs in 1498. May be, the dimensions of the Earth were not right, but Bacon began to consider the ocean not as an insurmountable barrier but as a route to the Indies.

**11. Dante and his Hell**
What about the problem of the Antipodes? Well, we find it in a poem, the Divina Commedia. Therefore, let us see how a poet, Dante Alighieri (1265–1321), viewed the world. His cosmos is illustrated in the Commedia, representing how the people imagine the world around the year 1300. In the Commedia and in other of his writings we find a lot of astronomical references [40]. He probably studied Aristotle in the commentaries of Thomas Aquinas, Pliny, and Al-Farghani. In Dante's Divina Commedia, the Hell is a conical cavity reaching to the centre of the earth. At the apex of the cone, there is Lucifer. After Dante and his guide Virgil have passed near Lucifer the bottom of the Hell, and continuing their journey, Dante looks back and sees Lucifer upside down. And Virgil explains that they have passed the center of the earth, which is pulling the weights (a clear statement on gravitation):

| ... tu passasti 'l punto<br>al qual si traggon d'ogne parte i pesi.<br>E se' or sotto l'emisperio giunto<br>ch'e` contraposto a quel che la gran secca<br>coverchia, e sotto 'l cui colmo consunto<br>fu l'uom che nacque e visse sanza pecca:<br>tu hai i piedi in su picciola spera<br>che l'altra faccia fa de la Giudecca. | …thou then didst pass the point to which<br>all gravities from every part are drawn.<br>And now thou art arrived beneath the hemisphere<br>opposed to that which canopies the great dry land,<br>and underneath whose summit was consumed the<br>Man, who without sin was born and lived; thou<br>hast thy feet upon a little sphere, which forms the<br>other face of the Judecca.<br>*The Inferno, Edited by Israel Gollancz, 1903* |
|---|---|

They commenced their ascent to the other side of the earth, toward the Antipodes, where they find the Purgatory, a conical hill, rising out of the ocean at a point diametrically opposite to Jerusalem. The terraces of the mount are their stairway to the earthly Paradise at the top and then to the celestial spheres. And here, during climbing, in the eleventh Canto of Purgatorio, there is an allusion to the precession of the equinoxes:

| Non è il mondan romore altro ch'un fiato<br>di vento, ch'or vien quinci e or vien quindi,<br>e muta nome perché muta lato.<br>Che voce avrai tu più, se vecchia scindi<br>da te la carne, che se fossi morto<br>anzi che tu lasciassi il 'pappo' e 'l 'dindi',<br>pria che passin mill'anni? ch'è più corto<br>spazio a l'etterno, ch'un muover di ciglia<br>al cerchio che più tardi in cielo è torto. | Naught is this mundane rumour but a breath<br>Of wind, that comes now this way and now that,<br>And changes name, because it changes side.<br>What fame shalt thou have more, if old peel off<br>From thee thy flesh, than if thou hadst been dead<br>Before thou left the 'pappo' and the 'dindi,'<br>Ere pass a thousand years? which is a shorter<br>Space to the eterne, than twinkling of an eye<br>Unto the circle that in heaven wheels slowest.<br>*Longfellow's Translation* |
|---|---|

During all his life, Dante was deeply interested in cosmology. In 1320, he delivered a lecture "Quaestio de Aqua et Terra", to discuss and refute the opinion that the water and the land of the earth were not the same sphere. During the Middle Age somebody believed that the earth was composed by two spheres: the land-sphere and the water-sphere, the centres of which were not coincident [3]. Dante knew very well the center of the earth was "the point to which all gravities from every part are drawn".

Let us end our short discussion on the medieval ideas about the form of the earth, with the words by which Dante ends the Hell. He is telling, after his long journey inside the earth, "we mounted up, he (Virgil) first and I second, so far that I distinguished through a round opening the beauteous things which Heaven bears; and thence we issued out, again to see the Stars".

In 1300, after a quite long journey of a thousand of years from Rome, the ancient philosophy had again a first role in the human knowledge. In the following, step by step as a climbing of a Purgatory, and following the works of some medieval philosophers like Grosseteste and Bacon, we arrive to Copernicus and Galileo and his Dialogue Concerning the Two Chief World Systems. Published on 1632 in Italian, Galileo compared the Copernican system with the Ptolemaic system. Even the life of Galileo was not peaceful: in September 1632, Galileo was ordered to come to Rome for a trial before the inquisitor. In 1633, the sentence of the Inquisition stated that Galileo was found "vehemently suspect of heresy"; he was forces to "abjure, curse and detest" his opinions. Galileo was sentenced to imprisonment, commuted to house arrest, which he remained under for the rest of his life. His Dialogue was banned and publication of any of his works forbidden, including any one he might write in the future.

On October 31, 1992, after 359 years, Pope John Paul II rehabilitated Galileo.

diametrically opposed to the origin of the divisions. In this way you will get five circles, the medium is equal to that determined by the circular line that was divided into parts. In the points where compass was pointed, place holes being sure that their centers coincide with the points of the compass, and into these holes fit tubes, which must be perfectly cylindrical and straight, so nothing prevents the view. It must fix them by means of a semicircle of iron to the sphere, so that they always maintain the same position. Hold the ball so that through the two poles you see the North Star, the Globe will be oriented, that line will be the meridian, and the others will represent the equator, the tropics and the polar circles.